# Stopping of 0.3 - 1.2 MeV/u protons and alpha particles in Si

M. Abdesselam[1], S. Ouichaoui[1,*], M. Azzouz[1], A.C. Chami[1] and M. Siad[2].

[1]USTHB, Faculté de Physique, B.P. 32, El Alia, 16111 Bab Ezzouar, Algiers, Algeria.
[2]CRNA/COMENA, 02 Bb Frantz Fanon BP Alger-gare, Algiers, Algeria.

**Abstract**

The stopping cross sections $\varepsilon(E)$ of silicon for protons and alpha particles have been measured over the velocity range (0.3-1.2) MeV/u from a $Si//SiO_2//Si$ (SIMOX) target using the Rutherford Backscattering Spectrometry (RBS) with special emphasis put on experimental aspects. A detection geometry coupling simultaneously two solid-state Si detectors placed at 165° and 150° relative to each side of the incident beam direction was used to measure the energies of the scattered ions and determine their energy losses within the stopping medium. In this way, the basic energy parameter, $E_x$, at the $Si/SiO_2$ interface for a given incident energy $E_0$ is the same for ions backscattered in the two directions off both the Si and O target elements, and systematic uncertainties in the $\varepsilon(E)$ data mainly originating from the target thickness are significantly minimized. A powerful computer code has been elaborated for extracting the relevant $\varepsilon(E)$ experimental data and the associated overall uncertainty that amounts to less than 3%. The measured $\varepsilon(E)$ data sets were found to be in fair agreement with H. Paul's compilation and with values calculated by the SRIM 06 computer code. In the case of $^4He^+$ ions, experimental data for the γ-effective charge parameter have been deduced via scaling the measured stopping cross sections to those for protons crossing the same target at the same velocity and compared to the predictions of the SRIM 06 code. It is found that the γ-parameter values generated by the latter code slightly deviate from experiment over the velocity region around the stopping cross section maximum where strong charge exchanges usually occur.

* Corresponding author, e-mail: souichaoui@gmail.com

# 1. Introduction

The stopping of swift charged particles in matter remains a subject of large experimental and theoretical concerns despite the large amount of stopping data accumulated giving rise to frequently improved tabulations over several decades [1, 2]. Several interaction mechanisms (elastic nuclear recoils, ionization and excitation of target atoms, excitation of the projectile while it moves within the target material, electron scattering and other processes) whose relative importance crucially depends on the projectile velocity govern the slowing down of such particles. Several stopping theories each being valid within a given velocity regime have been proposed earlier [3-5] for describing the basic energy loss processes. E.g., Lindhard's dielectric theory [5], using the Random Phase Approximation (RPA), has largely proved to be adequate for describing stopping cross section experimental data in the case of light particles such as protons and alpha particles. However, due to the complexity of the Barkas and Bloch corrections to the stopping cross section [6-11], the continuous charge exchange between incident ions and stopping target atoms, and the lack of experimental data at low ion velocities, the extension of these theories to moderately low energies has remained a subject of further investigations. Recently, a binary collision stopping theory (or binary collision approximation scheme, BCAS), applicable to the case of light charged particles has been developed by P. Sigmund and A. Schinner [12]. Extending Bohr's theory to higher energies, it also handles the complex low energy problems mentioned above. For projectile energies below and around the stopping cross section maximum, i.e., lying outside the energy interval of applicability of Bethe-Bloch theory, the ε ($E$) values can be calculated with adequately parameterizing the projectile charge in order to account for the charge state change within the solid mainly when the incident ion velocity becomes comparable to those of target atomic electrons. The complex effective charge dependence of the stopping cross section [12-15] is a longstanding problem still not fully elucidated at present. Of particular interest is the effective charge of swift heavy ions according to which the incident ion acts as a point charge whose value depends only on the ion velocity in first approximation. Even for alpha particles crossing solid targets, the effective charge parameter expression may be refined via convenient fits to experimental data. The most precise possible knowledge of the stopping cross section and straggling experimental data for light charged particles is of crucial importance for testing the validity of the above-mentioned stopping theories and underlying stopping concepts. Especially, the ion energy regions extending below and around the electronic stopping cross section maximum are of great importance in this respect. Besides, the accuracy of experimental data is also required for performing applications involving high-resolution ion-beam analyses. However, it must be noted that as the ion energy decreases down towards and below the stopping cross section maximum, the scatter in experimental data increases reaching sometimes more than 6%. This large dispersion has several causes: (i) systematic errors generated by target thickness, (ii) chemical and target density effects that may be strongly pronounced because of the larger contribution of valence electrons to the stopping process at low energies, (iii) the target texture influence which is increased at low energies due to larger critical channeling angles and (iv) the adopted experimental method that may influence the outcome results. Ziegler et al. [2] and H Paul [1] have reviewed the stopping cross section experimental data for protons in all elements and deduced corresponding semi-empirical fits. These tabulations exhibit large scatters in the measured data around the ε ($E$) function maximum, thus calling for additional careful stopping cross section measurements via high resolution methods within this energy region in order to reduce the observed large spreads.

Stopping cross section measurements using the Rutherford Backscattering Spectrometry (RBS) are less sensitive to the influence of target surface contaminations and non homogeneities in comparison to transmission experiments. The extraction of stopping cross section data from the RBS spectra is however more complex requiring mathematical approximations since the incident ions lose energy both along their ingoing and outgoing paths and during backscattering collisions in between. A mathematical effort is therefore need for extracting relevant experimental data. Although several methods are available for determining the stopping cross sections via the RBS technique [16-19], most of them are not enough accurate as required by many ion beam applications.

In undertaking the current study, we aimed at carefully determining the experimental stopping cross sections of silicon for $H^+$ and $^4He^+$ ions over the velocity range 0.25 - 1.0 MeV/u from a SIMOX target using the RBS technique in a suitable detection geometry. This method (see next section) greatly facilitates the determination of the ε ($E$) data, minimizing the associated systematic errors that mainly originate from the target thickness. The data for $^4He^+$ ions determined from the front edges of the O patterns in the RBS spectra allowed us to extend our previous numerical method [19] for exploring the lower ion energy region. We also aimed at extracting the effective charge parameter, γ, for the $^4He^+$ ions across the SIMOX target. The measured data are reported and compared to previous counterparts contained in Paul's compilation [1], and to values derived by Ziegler's SRIM 06 Monte Carlo simulation code [2].

## 2. Experimental

### 2.1. Set up and detection procedure

The experiments have been carried out at the CRNA Algiers 3.75-MV Van de Graff accelerator delivering light particles with energy $E > 0.8$ MeV and optimum energy resolution $\Delta E / E \approx 0.1\%$. Due to this energy limitation, $H_2^+$ and $H_3^+$ molecular ion beams have been also used to explore the lower energy region down to ~0.2 MeV for protons.

The energy calibration of the machine was made by measuring the γ-ray yields from the low energy resonances in the $^{19}$F (p, α γ) $^{16}$O reaction. H$^+$, H$_2^+$, H$_3^+$ and $^4$He$^+$ ion beams with average current of ~ 30 nA on the stopping target were used under high operation vacuum (mean pressure of about 10$^{-6}$ Torr) within the scattering chamber. The effectively estimated error on the incident beam energy was of about ±0.2%. A schematic diagram of the target arrangement and detection geometry is given in Fig. 1. To reduce the effects of surface contamination inside the scattering chamber, a cold trap was mounted near the target sample. It consisted of a cylinder coaxially surrounding the target, cooled down to liquid nitrogen temperature. This device strongly reduced the carbon contamination of the target surface and contributed to ensure a best assessment of the ion energy loss.

Each SIMOX target consisted of a SiO$_2$ layer produced in sandwich between two crystalline silicon layers. This material was chosen due to the high quality of the corresponding sample surface layer. Samples of different silicon layer thicknesses of 1.060, 1.750 and 2.220 × 10$^{18}$ atms/cm$^2$ were used in the experiments. The target sample thicknesses were determined through an accurate measurement of the energy losses of $^4$He$^+$ ions over the incident energy range $1.8 \leq E \leq 2.2$ MeV and performing a least-squares polynomial fit to the stopping cross section experimental data from Refs. [20, 21] involved in the ICRU 49 tabulation. Then, they were confirmed by scanning electron microscopy (SEM) imaging. Our fit to ε (E) experimental data over the preceding narrow projectile energy range was found in excellent agreement (to better than 0.6%) with the SRIM 06 code [2]. The error in target sample thickness was then estimated to be of about 1.5%, in average, and surely not to exceed 2%.

The sample surface non-homogeneity and non-uniformity were also investigated by alpha particle energy loss measurements, scanning each surface at several point positions at different ion beam energies. In general, the target foils used were found uniform to better than 1.5 %, this estimate generating the main experimental uncertainty in the measured ε (E) data (see below). To reduce the probable channeling effects generated by the material's texture, the Si target samples were double tilted to 5°- 6° with respect to both (θ, φ) angles in the spherical polar coordinate system where the incident beam direction and the normal to the target surface before orienting the latter were parallel along the Oz axis. Actually, the possible channeling effects were permanently controlled during the experimental runs using the RBS signals. A careful Nuclear Reaction Analysis scanning of the used targets before and after the stopping cross section measurements by means of 0.9 MeV deuterons showed no noticeable carbon contaminations.

Two solid state Si detectors located at 165° and 150° relative to each side of the beam direction were simultaneously used, associated to standard ORTEC detection electronics to measure the energy of the backscattered ions. The energy calibration of the detection system was performed by first recording the RBS spectrum from a $SiO_2$ target at the starting ion energy of 900 keV for $^3H^+$ and $^4He^+$, then using the front edges of the set of all recorded RBS spectra to determine the coefficients of the linear calibration curve. In this procedure, the ion energy losses within the dead-layers of the two detectors were taken into account: they were systematically corrected for in the calibration curve determination using the SRIM 06 code. This correction may be significant only for the Si stopping cross section values deduced from the oxygen front edge. In the current measurements, it amounted up to 2,5% of the total energy loss $\Delta E(J,\theta_i)$ (see Eq. 9 below) for low energy $^4He^+$ ions. The detection system energy resolution for the latter projectiles was estimated to be better then 12 keV (FWHM).

All the recorded RBS spectra for $H^+$ and $^4He^+$ ions were fitted by means of error functions in order to obtain precise values of the front edge positions and widths (FWHM) of the Si and O patterns, and determine the ion energy losses (see below). To achieve a good precision in the measured ε (E) data, a compromise was made to insure sufficient counting statistics within a reasonable counting time, without deteriorating the peak energy resolution: after several trials with using the same experimental set-up, RBS spectra were accumulated up to 1500-2000 counts per channel in the Si height surface layer such that both pulse pile-up and dead time problems were avoided.

## 2. 2. Data treatment and determination of energy losses

A typical energy spectrum for 2 MeV $^4$He$^+$ ions backscattered off the SIMOX target clearly showing the contributions of Si, SiO$_2$ and O layers is shown in Fig. 2 where the channels $C_0$, $C_1$, $C_2$ and $C_3$ corresponding to the energy distribution widths (FWHM) are the fundamental quantities for the energy loss determination. They were accurately determined using error function fits involving a chi-2 least squares minimization. The set of $C_0$ channels for all the RBS spectra were used in the detection system energy calibration.

As stated, several methods based on the RBS technique and applicable in some particular cases, each presenting manifest advantages as well as mathematical constraints, have been proposed [16-19] to extract the stopping cross section values. However, these methods become less reliable at low projectile energies in the case of thick and low Z-targets for which the kinematics factor, $k$, amounts to less than 0.7. To ovoid this limitation only to higher values of $k$ and incident energy $E$, we have used the following procedure. As suggested by Figs. (1, 2), we were interested in determining the whole width of a given energy loss distribution for outgoing ions, $\Delta E_{out}(E_0, \theta_i)$, measured at a detection angle $\theta_i$ (i =1, 2) by considering three distinct contributions:

(i) the energy loss, $\Delta E_{in}$, along the incoming ion path

(ii) the energy loss, $\Delta E_{out}(E_0, \theta_1)$, of outgoing ions scattered at angle $\theta_1$=150° and

(iii) the energy loss, $\Delta E_{out}(E_0, \theta_2)$, of outgoing ions scattered at angle $\theta_2$ = 165°.

These three quantities are deduced from the recorded RBS spectra by using the corresponding $C_0$, $C_1$ and $C_2$ characteristic channels and determining the projectile energy, $E_x$, just before the ion is scattered at the SiO$_2$ interface according to appropriate equations (see below). Two methods have been used to extract the stopping cross sections, $\varepsilon(E)$, from the so determined $\Delta E_{total}(E_0, \theta_i) = k(\theta_i) E_0(j) - E_{out}(j, \theta_i)$ total widths: the method described in details in Ref. [17] and the second one given below.

We start with a series of 2n RBS spectra (n from each detector) recorded for the same target at several (n) incident ion energies. The combined information resulting from the analysis of these spectra was then used to derive the ion energy losses during the incoming and outgoing paths. For each spectrum depicted by $j$ and corresponding to incident energy $E_0$ $(j)$, we search to deduce the ion energy $E_X$ $(j)$ just before the ion scattering off the $SiO_2$ layer, i.e., $\Delta E_{in}(E_0, \theta_{in}) = E_0(j) - E_X(j, \theta_{in})$, with $\theta_{in}$ being the angle between the incoming beam and the normal to the target (5-6°). To determine the energy loss upon scattering, $\Delta E_{out}(E_0, \theta_i) = k(\theta_i) E_x(j) - E_{out}(j, \theta_i)$, the energies of ions scattered both off Si and O elements present at the interface are considered. From the energy loss definition, this quantity can be related to the total stopping cross section, $\varepsilon(E)$, in function of the average incoming and outgoing energies, $\overline{E}_{in}$ and $\overline{E}_{out}^i$, for each of the two detection angles, $\theta_1 = 150°$ and $\theta_2 = 165°$, respectively. Thus, for a given incident energy, $E_0$ $(j)$, we extract five stopping cross section data: one for the alpha particle incoming path, two for alpha particles backscattered off silicon at the $Si/SiO_2$ interface, and two other ones for alpha particles backscattered off oxygen at the $Si/SiO_2$ interface. Then, one has the relations:

$$\Delta E_{in}(j, \theta_{in}) = \int_0^{e/\cos\theta_{in}} \varepsilon(E(x)) N dx = \varepsilon(\overline{E}_{in}, j) \times N\Delta X / \cos\theta_{in}. \quad (1)$$

and:

$$\Delta E_{out}(j, \theta_i) = \int_0^{e/\cos\theta_i} \varepsilon(E(x)) N dx = \varepsilon(\overline{E}_{out}^i, j) \times N\Delta X / \cos(\theta_i). \quad (2)$$

According to the latter equation, the target areal mass density, $\dfrac{e}{\cos\theta_{in}} = \dfrac{N\Delta X}{\cos\theta_{in}} = \int_{E_0}^{E_x} \dfrac{dE}{\varepsilon(E)}$, can be deduced as:

$$e = \cos(\theta_{in}) \int_{E_0}^{E_x} \dfrac{dE}{\varepsilon(E)} = \cos(\theta_1) \int_{k(\theta_1)E_x}^{E_1(j,\theta_1)} \dfrac{dE}{\varepsilon(E)} = \cos(\theta_2) \int_{k(\theta_2)E_x}^{E_2(j,\theta_2)} \dfrac{dE}{\varepsilon(E)}. \quad (3)$$

The average incoming and outgoing ion energies within the foil are given, respectively, by:

$$\overline{E}_{in}(j,\theta_{in}) = E_0 - \frac{\Delta E_{in}(j,\theta_{in})}{2} \quad (4)$$

and

$$\overline{E}^{i}_{out}(j,\theta_i) = E^{i}_{out}(j,\theta_i) + \frac{\Delta E_{out}(j,\theta_i)}{2}, \quad (5)$$

with

$$\Delta E_{in}(j,\theta_{in}) = E_0(j) - E_X(j,\theta_{in}) = \varepsilon(\overline{E}_{in}(j,\theta_{in})) \times Ndx \left|\sec\theta_{in}\right|, \quad (6)$$

$$\Delta E_{out}(j,\theta_i) = k(\theta_i)E_x(j) - E_{out}(j,\theta_i) = \varepsilon(\overline{E}_{out}(j,\theta_i)) \times Ndx \left|\sec\theta_i\right|. \quad (7)$$

The stopping cross sections before and after ion scattering, $\varepsilon(\overline{E}_{in},\theta_{in})$ and $\varepsilon(\overline{E}_{out},\theta_i)$, were obtained by iteration assuming straight trajectories and, hence, neglecting multiple ion scattering. They were expanded in power series [18] of energy, i. e.:

$$\varepsilon(E) = a_1 E^{-0.5} + a_2 E^{-0.25} + a_3 E^{0.25} + a_4 E^{0.5} + a_5 E^{0.75}. \quad (8)$$

A manifest advantage of this method resides in the analytical evaluation of the integral in Eq. (3). The numerical procedure consists in finding the energy $E_x$ (j) by solving Eqs. (3-7). Stable stopping cross section values are obtained after only a few iterations. For our configuration, the kinematics factor values for $^4$He$^+$ ions in Si are 0.568 and 0.585 for $\theta_1$=165° and $\theta_2$=150°, respectively. Being conclusive in the case of $^4$He$^+$ ions backscattered off oxygen, this method is then extended to proton backscattering characterized by a lower kinematics factor.

Finally, the energy loss data in silicon have been determined following the above procedure both for H$^+$ and $^4$He$^+$ ions over the explored velocity ranges of 0.25 - 1.2 Mev/u and 0.075 - 0.75 MeV/u, respectively. Corresponding best fit error functions have been generated to determine with high accuracy the $C_0(j,\theta_i)$, $C_1(j,\theta_i)$ and $C_2(j,\theta_i)$ (see Fig. 2) channels corresponding to energy distribution front edges, from which the total energy loss values were obtained using the relation:

$$\Delta E(J,\theta_i) = a\left[C_0(j,\theta_i) - C_1(j,\theta_i)\right] = k(\theta_i)\Delta E_{in}(j,\theta_{in}) + \Delta E_{out}(j,\theta_i). \quad (9)$$

where *a* represents the calibration constant. Then, the Si stopping cross section experimental data are extracted from Eqs. (1, 2).

## 2.3. Error analysis

The thickness uniformity of the films was checked by comparing the widths and yields of the energy loss distributions, scanning the film surfaces at different beam spot impacts. The film surfaces were found to be uniform to better than the precision afforded by the RBS technique (relative uncertainty < 0.5%) that is included in the errors in target thickness. The relative uncertainty in the energy calibration of the experimental set up was estimated to be lower than 0.4 %,

Several error sources were considered in our stopping cross section measurements. The uncertainties (assumed to be uncorrelated) essentially arose from the determination of the target thickness, the energy loss distribution widths and the calculation procedure used to extract the fundamental parameter, $E_X(j, \theta_{in})$. The simultaneous use of two detectors considerably reduced the systematic uncertainties in the target thickness determination estimated to be of the order of that from the SRIM 06 code in the Si stopping cross section for 2 MeV $^4$He$^+$ ions, i.e., ~1.5%, in average, and at most 2% as stated above (see Subsection 2.1). The nuclear reaction analysis of the samples using a 0.9 MeV deuteron beam showed that they suffered no significant C and O contaminations. The uncertainty in the measured energy-loss distribution widths and that introduced by the data analysis procedure and the approximations made in our calculation code were estimated to be lower than 2%. Notice that low energy data are affected by an additional uncertainty due to the significantly reduced kinematics factor values and to molecular beam effects [22]. Finally, the overall uncertainty in the measured stopping cross sections were estimated to be of about 2.5 % when deduced for ions backscattered off Si atoms at the interface, and to about 4 % relative to oxygen diffusing atoms because of the cited preceding effects.

## 3. Results and discussion:

As already stated, energy loss values for H$^+$ or $^4$He$^+$ ions backscattered off SIMOX targets were simultaneously determined at angles of 165° and 150°, thus substantially reducing systematic errors.

Our stopping cross section results are reported in Figs. (3, 4) where they are compared to previously measured data from H. Paul's compilation [1] and to semi-empirical values derived by the SRIM06 [2] computer code. Notice that the stopping cross sections of Si generated by the TRIM 85 code [2] show substantial inconsistencies with the current experimental data for both $H^+$ and $^4He^+$ ions, as expected. This point has been discussed previously by several groups (e.g., [23] and Refs. therein), which has led to update the Si stopping cross section data in Ziegler's SRIM 2003 version mainly on the basis of the ε (E) data measured by G. Konac et al. [24].

### 3. 1. Stopping cross section data:

### 3. 1. 1. $H^+$ ions in Si:

As shown by Fig. 3, our stopping cross section data for $H^+$ ions in Si are consistent with those reported in Ref. [24], by D. Niemann et al. [25], and with values derived by the SRIM 06 code [2] while the data from Ref. [26] lie slightly below our data. As can be seen in this figure, our data for E > 0.5 MeV are in good agreement to within 1% with those reported in the literature.

### 3. 1. 2. $^4He^+$ ions in Si:

Our ε (E) data for $^4He^+$ ions in Si are compared in Fig. 4 to previous ones from the literature and to values calculated by the SRIM 06 code. As can be seen, they are in excellent agreement with those from Refs. [24, 20, 21] showing an overlapping of error bars, and with values calculated by the SRIM 06 code over the whole explored ion energy range. One can note that over the energy range 0.8 - 2.0 MeV, the data from Refs. [20] and [21] systematically lie to within ~3 % above and below our data, respectively. Besides, Fig. 4 indicates that around the stopping cross section maximum, the data from Refs. [27, 20] are systematically higher than our values by ~ 4–8% and 1-3%, respectively. …….In the energy range 0.3-0.5 MeV, a net scatter is observed in our data which, however, remain consistent with the SRIM 06 code calculation. This can be attributed to particular difficulties in determining the ion energy losses from the energy loss distribution front edges for $^4He^+$ ions back-scattered off oxygen atoms at the Si/SiO2 interface. These difficulties are due to several effects taking place at low energies: (i) while the backscattering cross section rapidly increases with decreasing incident ion energy, (ii) multiple scattering become significant, and (iii) the kinematics factor associated with oxygen atoms is low ($k = 0.366$). As a result, the RBS signal front edges undergo deterioration and the handling of our computer code for treating the RBS spectra is more complicated. This spreading in the low energy data induces an error of the order of 4-5% in the experimental stopping cross section. It gradually reduces with increasing $^4He^+$ ion energy, practically vanishing above $E = 0.5$ MeV. One must remark, however, that the above effects do not noticeably affect the stopping cross section determination for $^4He^+$ ions backscattered off Si atoms over the energy range explored.

### 3. 2. Effective charge parameter data for $^4He^+$ ions in Si:

The indirect method for deriving the charge state of swift ions inside solid materials from energy loss measurements has led to the concept of the effective charge parameter supposed to yield information on the slowing down ion's equilibrium charge state. According to Bohr's criterion, the projectile (of initial charge $ze$) is supposed to be fully stripped of all its electrons whose orbital velocities are lower than the ion velocity v. The idea was to consider the ion stopping only within the first order perturbation theory by assuming $\varepsilon(E)$ to be proportional to $z^2$ for fully stripped ions. Then, a $z^2$ - scaling law is usually adopted in order to predict the electronic stopping cross section of the target atoms (charge $Z_2e$) for heavy ions, $\varepsilon_e(z,v,Z_2)$, relative to that of protons, $\varepsilon_e(p,v,Z_2)$, crossing the same target with the same velocity, i.e.:

$$\varepsilon_e(z,v,Z_2) = (\gamma z)^2 \, \varepsilon_e(p,v,Z_2), \tag{10}$$

where γ is the effective charge parameter thus given by:

$$\gamma = \frac{1}{z}\sqrt{\frac{\varepsilon_e(z_{He},v,Z_2)}{\varepsilon_e(p,v,Z_2)}}. \tag{11}$$

To our knowledge, no theory is presently available for calculating this parameter whose justification mainly comes from experiment. The usual form of $\gamma$ for the He$^+$ ions is given by [15]:

$$\gamma_{He} = 1 - \xi \exp(-\lambda\, v_r). \qquad (12)$$

In this equation, $v_r = v/v_0\, z^{2/3}$ denotes the ion reduced velocity (with the Bohr velocity given by $v_0 = c/137$), $\xi$ and $\lambda$ are adjustable parameters, and the fractional effective charge for protons is assumed to be always unity.

The current stopping cross section measurements over the velocity range 1.2 to 3.5 $v_r$ for H$^+$ and $^4$He$^+$ ions in Si gives us the opportunity to analyze the results in terms of the effective charge parameter $\gamma$. Then, expression (12) has been fit to the measured data from this work as well as to data reported in the literature [1], leading to the following analytical expression:

$$\gamma_{He} = 1 - 1.40 \exp(-1.85\, v_r) \quad, \qquad (13)$$

represented by the solid curve in Fig. 5 where the effective charge parameter values are plotted versus the relative velocity for $^4$He$^+$ ions in Si. A very good account of experimental data by Eq. (12) is thus obtained over the whole velocity range, except for very low values of $v_r$. The $\gamma$ parameter values derived by the SRIM 06 [2] computer code under the $z^2$- dependence assumption are given in Fig. 5. As can be seen, they show to deviate slightly from experimental data over the relative velocity range $v_r = 1 - 2.7$ corresponding to the stopping cross section maximum (see also Figs. 3-4). As expected, at the highest velocity values, $v_r > 4$, corresponding to fully stripped helium ions ($^4$He$^{++}$), both calculations obviously give $\gamma \sim 1$.

**Conclusions**

The use of SIMOX targets has allowed us to measure with a good precision the Si stopping cross sections for protons and $^4$He$^+$ ions over the velocity range 0.25 - 1.0 MeV/amu. The experimental set up used based on simultaneously coupling two solid state Si detectors in a RBS configuration has facilitated the determination of the projectile energy, $E_x$, at the Si/SiO$_2$ interface and allowed to considerably reduce systematic errors in the measured data mainly originating from the target thicknesses. The stopping cross section values were deduced by making use of our previous computation method [19]. We therefore confirm the efficient extension of applying the RBS technique to the case of protons and $^4He^+$ ions crossing low-Z materials, i.e., in case of low kinematics factor values. The measured $\varepsilon(E)$ data are found to be in a general good agreement with those reported in H. Paul's compilation [1], and are also very consistent with values calculated by the SRIM 06 code. A set of about 500 stopping cross section values have been checked under the $z^2$ - scaling law over the velocity range (0.7- 3) $v_r$ in order to extract the effective charge parameter, $\gamma$, for helium ions backscattered off Si. The determined effective charge parameter data are satisfactorily described by our exponential expression (13) consistently with the suggestion of L. C. Northcliffe and by H.D. Betz [15]. Besides, the $\gamma$ values calculated by the SRIM 06 code show slight deviations from experimental data over the ion relative velocity region around the stopping cross section maximum. It thus turns out that cautions must be taken in using the latter code for calculating the effective charge of charged ions in solid materials in the ion velocity region of strong charge exchanges.

**Acknowledgement:**

We thank the CNRA/ DTN technical staff for their cooperation, especially A. Midouni whose kind assistance during the experiments was highly appreciated. The authors are also very grateful to Dr. J.J. Grob from the InESS of Strasbourg for performing the SIMOX targets.

**References:**

[1] Helmut Paul, Institut für Experimental physik, Universität of Linz A-4040, Linz, Austria; http://www.exphys.uni-linz.ac.at/Stopping/.

[2] J. F. Ziegler, J. P. Biersack and U. Littmark. The stopping and Range of Ions in Solids (Pergamon Press, New Yorh, 1985); J. F. Ziegler, SRIM 06 at http://www.srim.org/index.htm.

[3] A. Bohr, Mat. Fys. Medd. Dan. Vid. Selsk., **18**, No. 8 ( 1948).

[4] H.A. Bethe, Ann. Phys., 5 (1930) 325., Phys. Rev., **89** (1953) 1256.

[5] J. Lindhard, Mat. Fys. Medd. Dan. Vid. Selsk., **28**, No.8 (1954).

[6] W. H. Barkas and M. J. Berger, in Penetration of Charged Particles in Matter, 1st printing, ed. U. Fano, (National) Academy of Sciences--National Research Council, Washington, D. C., 1964), Publ. 1133.

[7] F. Bloch, Ann. Phys. (Leipzig) **16** (1933) 287 .

[8] S. P. Ahlen, Rev. Mod. Phys. **52** (1980) 121.

[9] E. Bonderup, Mat. Fys. Medd. Dan. Vid. Selsk., **35** No. 17 (1967).

[10] J. Lindhard, Nucl. Instr. and Meth. **132** (1976) 1.

[11] B.S. Yarlagadda, J.E. Robinson and W. Brandt, Phys. Rev. **B** 17 (1978) 3473.

[12] P Sigmund and A. Schinner, Nucl. Instr. and Meth. **B** 193 (2002) 49-55; Europ. Phys. J. **D** 12 (2000) 425.

[13] W. Brandt and M. Kitagawa, Phys. Rev. **B** 25 (1982) 5631.

[14] M. D. Brown and C. D. Moak, Phys. Rev. **B** 6 (1972) 90.

[15] C. Northcliffe, Phys. Rev. 120 (1960) 1744; see also. H.D. Betz, Rev. Mod. Phys. **44** (1972) 465.

[16] W. K. Chu, J. W. Mayer and M.A. Nicole (eds). Backscattering Spectrometry (Adademic Press. New York. 1978 pp. 276-287; see also: W. D. Warters, Thesis, California Institute of Technology (1953).

[17] J. P. Stoquert, M. Abdesselam, H. Beaumevielle, Y. Boudouma and J.C. Oberlin, Nucl. Instr. and Meth. **194** (1982) 51; C. Obedin, A. Amokrane. H. Beaumevielle and J. P. Stoquert, J. Phys. **43** (1982) 485.

[18] C. Eppacher and D. Semrad, Nucl. Instr. and Meth. **B** 35 (1988) 109.

[19] M. Abdesselam, J. P. Stoquert, M. Hage-Ali, J.J. Grob and P. Siffert, Nucl. Instr. and Meth. **B** 73 (1993) 115.

[20] N.P. Barradas, C.Jeynes, R.P.Webb and E.Wendler, Nucl. Instr. and Meth. **B** 194 (2002) 15.

[21] D.C. Santry and R.D. Werner, Nucl. Instr. and Meth. **188** (1981) 211; Nucl. Instr. and Meth. **185** (1981) 517.

[22] W. Brandt, A. Ratkowski and R;H; Ritchie, Phys. Rev. lett. **33** (1974) 135.

[23] G. Boudreault, C. Jeynes, E. Wendler, A. Nejim, R. P. Webband U. Wätjen, Surface and Interface Analysis, Vol. **33** (2002) 478-486.

[24] G. Konac, S. Kalbitzer, Ch. Klatt, D. Niemann, R. Stoll, Nucl. Instr. and Meth. **B** 136-138 (1998) 159.

[25] D. Niemann, G.Konac, S.Kalbitzer, Nucl. Instr. and Meth. **B** 118 (1996) 11; Nucl. Instr. and Meth. **B** 124 (1997) 646.

[26] A. Ikeda, K. Sumimoto, T. Nishioka, Y. Kido, Nucl. Instr. and Meth. **B** 115 (1996) 34.

[27] C. Eppacher, PhD Thesis, Univ. of Linz, Austria, Schriften der Johannes-Kepler Universität Linz, Universitätsverlag Rudolf Trauner (1995).

[28] M. Fama, G.H. Lantschner, J.C. Eckardt, N.R. Arista, J.E. Gayone, E. Sanchez, F. Lovey Nucl. Instr. and Meth. **B** 193 (2002) 91.

**Figures :**

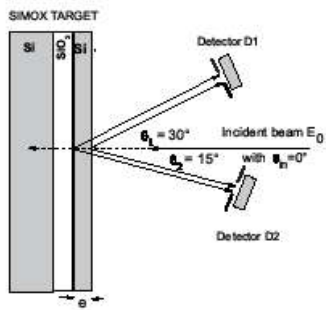

Figure 1

Fig.1: Schematic diagram of the experimental geometry:

- $E_0$ is the primary beam energy,

- $\theta_1$ and $\theta_2$ are the used backsattering angles ($\theta_i$),

- $k(\theta_i)$ is the kinematic factor of a projectile the bacscattered at $\theta_i$,

- $E_1(\theta_1)$ and $E_2(\theta_2)$ are the energies of the backscattered ions at angles $\theta_1$ and $\theta_2$ respectively,

- $e = N\Delta x$ is the area density of the Si target,

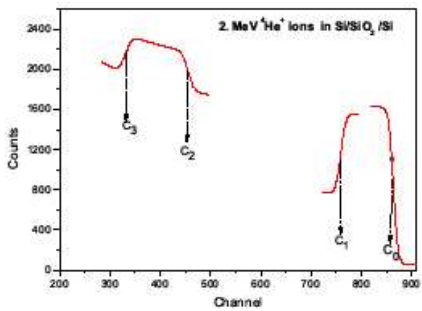

Fig. 2: Typical RBS spectrum for 2. MeV $^4$He$^+$ ions backscattered off a Si/SiO$_2$/Si target; channels $C_0$ and $C_1$ correspond to the energies of $^4$He$^+$ ions backscattered at the front and rear of the Si surface layer, respectively, while channels $C_2$ and $C_3$ refer to $^4$He$^+$ ions backscattered at the front and rear of the SiO$_2$ sandwich layer.

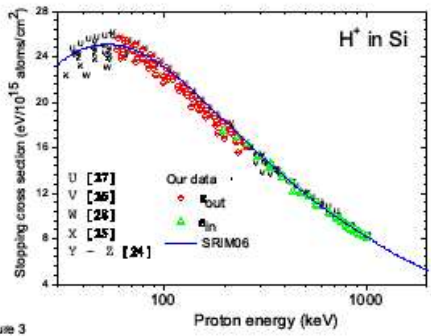

Fig. 3: Energy dependence of the electronic stopping cross section of Si for protons. Our measured data are compared to those reported in the H. Paul compilation (1). The solid curve represents values calculated by the SRIM 06 code.

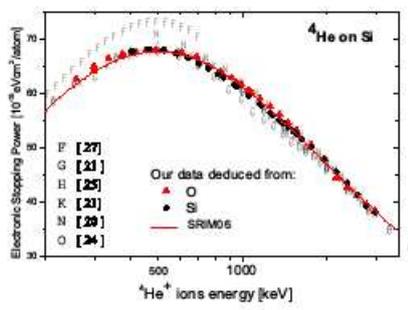

Fig. 4: Same as Fig.3 for $^4$He$^+$ ions.

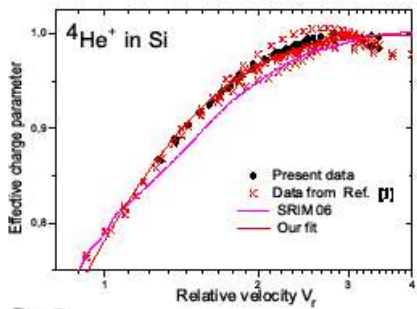

Figure 5. Effective charge parameter $\gamma^*$ for $^4He^+$ ions in Si derived by Eq. (12) in function of the ion reduced velocity, $V_r$. Our values, fitted by Eq. (13), are compared to Brandt-Kitagawa (20) and SRIM06 code calculations.